\documentclass[12pt]{article}
\usepackage{slashed,amsmath,amssymb,enumitem}
\usepackage[papersize={8.5in,11in}]{geometry}
\begin{document}
\title{Ultraviolet Complete Quantum Field Theory and Gauge Invariance}
\author{J. W. Moffat\\~\\
Perimeter Institute for Theoretical Physics,\\ Waterloo, Ontario N2L 2Y5, Canada\\
and\\
Department of Physics, University of Waterloo,\\ Waterloo, Ontario N2L 3G1, Canada}
\date{\today}
\maketitle
\begin{abstract}%
The problem of gauge invariance in an ultraviolet complete quantum field theory (QFT) with nonlocal interactions is investigated. For local fields that couple through a nonlocal interaction, it is demonstrated that the quantum electrodynamic (QED) sector of the electroweak (EW) model without a Higgs particle is gauge invariant. The non-Abelian QFT with massless gluons and nonlocal interactions is shown to be gauge invariant, and a perturbative S-matrix formalism in the interaction representation is investigated. The finite self-energy of a fermion and vacuum polarization are derived to first order. The K\"all\'en-Lehmann representation for the finite QED theory for local Heisenberg field operators is considered and the Coulomb potential between two charges is shown to be non-singular at $r=0$. The high energy limit of electromagnetic form factors is investigated in relation to the finite charge renormalization constant $Z_3^{-1}$. The magnitudes of the constants $Z_1,Z_2$ and $Z_3$ for charge and mass renormalization are derived without divergences and shown to be non-perturbatively finite. The proton-proton collision cross section production of lepton pairs through the Drell-Yan process is investigated and the prediction for the cross section at high energies can be tested at the LHC.
\end{abstract}


\section{Introduction}

An electroweak (EW) theory has been proposed in which nonlocal interactions of the quarks, leptons the $W$, $Z$, and massless photons and gluons are consistent with all presently viable EW data and unitarity~\cite{Moffat,Moffat2}. The theory is made finite by postulating that the coupling constants $e$, $g$ and $g'$ are functions of the center-of-mass energy $\sqrt{s}$:
\begin{equation}
{\bar e}(s)=e{\cal E}(s),\quad {\bar g}(s)=g{\cal E}(s)\quad {\bar g}'(s)=g'{\cal E}(s),
\end{equation}
where ${\cal E}(s)$ is an entire function of $s$ analytic (holomorphic) in the complex plane, except for singularities at infinity.

In earlier work on nonlocal quantum electrodynamics (QED)~\cite{Moffat3}, the fermion and photon fields were made nonlocal and the fermion and boson propagators were modified by an entire function. The interaction terms were smeared out by nonlocal entire function factors, resulting in a violation of gauge invariance at every order of perturbation theory due to the breaking up of the covariant derivative operator. The gauge invariance was restored by adding compensating contributions at every order of the perturbation theory expansion. This was proved to be valid for QED to all orders of perturbation theory, and the method was extended to non-Abelian gauge theory~\cite{Moffat3} as well as for gravity~\cite{Moffat4}.

In the UV complete EW theory the fermions and bosons are never massless and there is no phase in the universe in which $SU(2)\times U(1)$ is symmetric~\cite{Moffat}. The massless photon is described by a $U_{\rm EM}(1)$ gauge invariant sector and the massless gluon is part of the gauge invariant colored quark QCD described by the $SU_C(3)$ color group.

In the following, we will concentrate on the predictions of QED processes in our UV complete theory. The high energy behavior of electromagnetic form factors is analyzed and the non-perturbative results for the magnitudes of the renormalization constants $Z_1, Z_2$ and $Z_3$ are studied, using a K\"all\'en-Lehmann spectral representation. Finally, predictions for high energy proton-proton collisions at the LHC for the Drell-Yan process $q\bar{q}\rightarrow V^*\rightarrow\ell\bar\ell$ are derived that can be tested at the LHC.

\section{Nonlocal Gauge Invariant QED}

We choose the following notation: The Minkowski metric has the signature, $\eta_{\mu\nu}=(+1, -1, -1, -1)$, $\gamma_\mu\gamma_\nu+\gamma_\nu\gamma_\mu=2\eta_{\mu\nu}$ and $\slashed p=\gamma^\mu p_\mu$. Let us begin by investigating the standard theory of QED with local interactions. Consider the local gauge transformation for the Dirac field $\psi$:
\begin{equation}
\label{phaserotation}
\psi(x)\rightarrow \psi'(x)=\exp(i\theta(x))\psi(x),
\end{equation}
where $\theta(x)$ is the position dependent phase. We obtain the gradient:
\begin{equation}
\partial_\mu\psi(x)\rightarrow\partial_\mu\psi'(x)=\exp(i\theta(x))[\partial_\mu\psi(x)+i\partial_\mu\theta(x)\psi(x)].
\end{equation}
The additional gradient contribution spoils the local gauge invariance. To obtain gauge invariance, we replace $\partial_\mu$ by the covariant derivative
\begin{equation}
\label{covariantder}
D_\mu=\partial_\mu+ieA_\mu,
\end{equation}
where $A_\mu$ denotes the electromagnetic 4-potential and $e$ is the charge of the particle in natural units. The field $A_\mu$ transforms under the phase rotation (\ref{phaserotation}) as
\begin{equation}
\label{potentialtransf}
A_\mu(x)\rightarrow A_\mu'(x)=A_\mu(x)-\frac{1}{e}\partial_\mu\theta(x).
\end{equation}
We now find that under the local gauge transformation
\begin{equation}
\label{gaugecondition}
D_\mu(x)\psi(x)\rightarrow \exp(i\theta(x))D_\mu(x)\psi(x).
\end{equation}

The QED Lagrangian is given by
\begin{equation}
{\cal L}_{\rm QED}=-{\bar\psi}(x)(i{\slashed D}(x)-m)\psi(x)-\frac{1}{4}F^{\mu\nu}(x)F_{\mu\nu}(x),
\end{equation}
where
\begin{equation}
\label{Ftensor}
F_{\mu\nu}=\partial_\mu A_\nu-\partial_\nu A_\mu.
\end{equation}
The QED Lagrangian is invariant under the local gauge transformations (\ref{phaserotation}) and (\ref{potentialtransf}). We have
\begin{equation}
{\cal L}_{\rm QED}={\cal L}_{0D}-\frac{1}{4}F^{\mu\nu}F_{\mu\nu}+{\cal L}_I,
\end{equation}
where
\begin{equation}
{\cal L}_{0D}=-{\bar\psi}(i\slashed\partial-m)\psi
\end{equation}
and
\begin{equation}
{\cal L}_I=e{\bar\psi}(x)\gamma^\mu\psi(x)A_\mu(x)=J^\mu(x) A_\mu(x).
\end{equation}

To obtain a fully finite nonlocal QED free of divergences, we must break the manifest covariant derivative gauge invariance by nonlocalizing only the interaction term:
\begin{equation}
\label{nonlocalint}
e{\bar\psi}(x)\gamma^\mu\psi(x)A_\mu(x)\rightarrow {\bar e}(x){\bar\psi}(x)\gamma^\mu\psi(x)A_\mu(x),
\end{equation}
where ${\bar e}(x)=e{\cal E}(x)$ and ${\cal E}(x)$ is an entire function of $\Box=\partial^\mu\partial_\mu$. Eq.(\ref{nonlocalint}) can be rewritten in the form
\begin{equation}
e{\bar\psi}(x)\gamma^\mu\psi(x)A_\mu(x)\rightarrow e{\bar\Psi}(x)\gamma^\mu\Psi(x)A_\mu(x),
\end{equation}
where
\begin{equation}
\Psi(x)={\cal E}(x)\psi(x).
\end{equation}

That the apparent lack of a gauge symmetry for nonlocal QFTs cannot apply to all nonlocal gauge theories follows from the existence of string theory and string field theory~\cite{Witten,Witten2,Zwiebach}, which possess a nonlocal gauge symmetry. The important idea is to extend the notion of ``gauge invariance'' to nonlocal transformation laws. The key to the success of gauge symmetry in QFT is the decoupling of unphysical vector field and tensor field modes, while maintaining Poincar\'e invariance. A symmetry that succeeds in this is acceptable, {\it and the transformation rule need not be local}. The transformation rule contains two parts: 1) a local inhomogeneous term, which preserves the local quadratic part of the action, and a nonlocal inhomogeneous part, which generates a variation of the free action that cancels the inhomogeneous variation of the nonlocal interaction. The existence of a suitable nonlocal gauge invariance is not limited to string theory. Any local QFT can be generalized to a finite, nonlocal gauge theory
with a nonlocal gauge symmetry that maintains unitarity and Poincar\'e invariance. The replacement (\ref{nonlocalint}) is invariant at order $e$ under the transformations:
\begin{equation}
\delta A_\mu=-\partial_\mu\theta,
\end{equation}
and
\begin{equation}
\delta\psi=ie{\cal E}\Theta\Psi,
\end{equation}
where $\Theta={\cal E}\theta$ and the explicit differential operator ${\cal E}$ is understood to act on everything to its right. Invariance is lost at order $e^2$, and we cannot modify the transformation law to recover it. We must then add higher-order terms, both to the action and the symmetry~\cite{Moffat3}. In contrast to string theory our UV complete nonlocal QFT avoids the need for higher-dimensions; the theory is formulated in four-dimensional spacetime.

In refs.\cite{Moffat3,Moffat}, it was shown by means of a general transformation rule how gauge invariance implies current conservation and decoupling of unphysical modes. Let us consider the QED action of the form:
\begin{equation}
\label{QEDaction}
S_{QED}=-\int d^4x \biggl[\frac{1}{4}F^{\mu\nu}F_{\mu\nu}+\bar\psi(i\slashed\partial-m)\psi\biggr]-\int d^4xd^4y\bar\psi(x){\cal V}[eA](x,y)\psi(y),
\end{equation}
where $F_{\mu\nu}$ is given by (\ref{Ftensor}). The vertex operator ${\cal V}[eA]$ is in general a spinorial matrix and is formed from entire functions. It can be expanded in a power series ${\cal V}\sim eA+(eA)^2+\cdots$. We ignore the possibility of pure photon and multifermion interactions, for they cannot be used to restore gauge invariance and decoupling. Let us suppose that the interaction is invariant under the transformations:
\begin{equation}
\delta A_\mu(x)=-\partial_\mu\theta(x),
\end{equation}
and
\begin{equation}
\label{Toperator}
\delta\psi(x)=ie\int d^4yd^4z{\cal T}[eA](x,y,z)\theta(y)\psi(z).
\end{equation}
The operator ${\cal T}\sim 1+eA+\cdots$ is a spinorial matrix and a functional of the vector potential $A_\mu$.

Let us recapitulate the explicit proof of gauge invariance in our nonlocal QED~\cite{Moffat3}. We introduce the operator ${\cal O}$:
\begin{equation}
{\cal O}\equiv \frac{({\cal E})^2-1}{\Box+m^2},
\end{equation}
where ${\cal O} $ is an entire function of $\Box$. By utilizing the operator ${\cal O}$, we can determine the four-point interaction:
\begin{equation}
{\cal L}_2=-e^2{\bar\Psi}\slashed{A}(i\slashed\partial+m){\cal O}\slashed{A}\Psi.
\end{equation}
The Compton scattering amplitude calculated to order ${\cal L}_{0+1+2}$ is the same as in standard local QED. The two parts in the calculation determined by the decomposition of the operator ${\cal O}$: ${\cal E}^2/(\Box+m^2)$ and $-1/(\Box+m^2)$ cancel to give the usual result in the physical channel. This can be extended to higher-order interactions
\begin{equation}
{\cal L}_n=-(-e)^n{\bar\Psi}\slashed{A}[(i\slashed\partial+m){\cal O}\slashed{A}]^{n-1}\Psi.
\end{equation}
A summation of this result yields the total Lagrangian:
\begin{equation}
{\cal L}_{QED}=-\frac{1}{4}F_{\mu\nu}F^{\mu\nu}-\bar{\psi}(i\slashed{\partial}-m)\psi
+e\bar\Psi\slashed{A}[1+e(i\slashed{\partial}+m){\cal O}\slashed{A}]^{-1}\Psi.
\end{equation}
Because the Compton tree graphs are the same as those of local QED, the decoupling of unphysical modes is manifest. Only Feynman diagrams containing internal photon or fermion lines differ from local QED, through an enhancement by the regularizing entire function ${\cal E}$.

The nonlocal action is now invariant at each order under the transformation
\begin{equation}
\delta A_\mu=-\partial_\mu\theta,
\end{equation}
and
\begin{equation}
\delta_n\psi=-i(-e)^n{\cal E}\Theta[(i\slashed\partial+m){\cal O}\slashed{A}]^{n-1}\Psi.
\end{equation}
By summation we obtain
\begin{equation}
\delta A_\mu=-\partial_\mu \theta,
\end{equation}
and
\begin{equation}
\label{gaugetransf}
\delta\psi=ie{\cal E}\Theta[1+e(i\slashed\partial+m){\cal O}\slashed{A}]^{-1}\Psi.
\end{equation}
We have\footnote{For a detailed proof, see ref.\cite{Moffat3}, Sect. III, Eqs.(3.11a)-(3.11e)}
\begin{eqnarray}
\delta_0{\cal L}_n&=&-\delta_n{\cal L}_0-\delta_{n-1}{\cal L}_1-...-\delta_1{\cal L}_{n-1}\nonumber\\
&=&\bar\psi(i\slashed\partial-m)\delta_n\psi+(-e)^{n-1}\overline{\delta_1\Psi}
\slashed{A}[(i\slashed\partial+m){\cal O}\slashed{A}]^{n-2}\Psi+...
+\overline{\delta_n\psi}(i\slashed\partial+m)\psi.
\end{eqnarray}
It follows from this that $\delta{\cal L}=0$ to order $e^n$. From this result, we can infer the form of the operator ${\cal T}[eA]$:
\begin{equation}
{\cal T}[eA](x,y,z)={\cal E}[\delta^4(x-y)][1+e(i\slashed\partial+m){\cal O}\slashed{A}]^{-1}{\cal E}\delta^4(x-y).
\end{equation}

\section{Non-Abelian Gauge Theory and Nonlocal Interactions}

The formulation of non-Abelian (Yang-Mills) gauge theory proceeds as in the Abelian $U_{\rm EM}(1)$ case. Consider the local gauge transformation for the Dirac field operator:
\begin{equation}
\psi(x)\rightarrow\psi'(x)=U(x)\psi(x),
\end{equation}
where
\begin{equation}
U(x)=\exp\biggl(\frac{i}{2}T^a\theta^a(x)\biggr),
\end{equation}
and $T^a$ are the generators of the unitary group $SU(N)$. Then, the gradient transforms as
\begin{equation}
\partial_\mu\psi(x)\rightarrow U(x)(\partial_\mu\psi(x))+(\partial_\mu U(x))\psi(x).
\end{equation}
We ensure the gauge invariance of the theory by introducing the covariant derivative operator:
\begin{equation}
D_\mu(x)= \partial_\mu+igB_\mu(x),
\end{equation}
where $B_\mu=T^aB_\mu^a$. The $B_\mu$ satisfies the gauge transformation:
\begin{equation}
B'_\mu(x)=U(x)\biggl[B_\mu(x)+\frac{i}{g}U^{-1}(x)(\partial_\mu U(x))\biggr]U^{-1}(x).
\end{equation}

The gauge invariant Lagrangian is given by
\begin{equation}
{\cal L}_{NA}=-\bar\psi(x)(i\slashed{D}(x)-m)\psi(x)-\frac{1}{4}B^{\mu\nu}(x)B_{\mu\nu}(x),
\end{equation}
where
\begin{equation}
B_{\mu\nu}(x)=\frac{1}{ig}[D_\mu(x),D_\nu(x)]=\partial_\mu B_\nu(x)-\partial_\nu B_\mu(x)+ig[B_\mu(x),B_\nu(x)].
\end{equation}
We can write this equation in the form:
\begin{equation}
B^a_{\mu\nu}(x)=\partial_\mu B^a_\nu(x)-\partial_\nu B^a_\mu(x)+gf_{abc}B^b_\mu(x)B^c_\nu(x),
\end{equation}
where $f_{abc}$ are the structure constants of the group $SU(N)$.

A nonlocal gauge invariant, Poincar\'e invariant and UV complete non-Abelian Yang-Mills theory can be formulated describing QCD with colored quarks and an octet of massless gluons~\cite{Moffat3}. We begin by nonlocalizing the Yang-Mills gauge field:
\begin{equation}
{\cal B}^{a}_\mu={\cal E}(x,\Lambda_{NA})B^a_\mu,
\end{equation}
where $\Lambda_{NA}$ is the non-Abelian energy scale. We note that in contrast to $U(1)$ QED, in order to guarantee that all loop diagrams are finite, we must make the gauge field $B^a_\mu$ nonlocal, because of the massless gluon gauge field interactions. This nonlocalization of $B^a_\mu$ destroys the gauge invariance at order $g$ and higher. We can restore gauge invariance at order $g^2$ by the nonlocal gauge transformation:
\begin{equation}
\delta B^a_\mu=-\partial_\mu\theta^a+gf_{abc}{\cal E}{\cal B}^b_\mu\Theta^c,
\end{equation}
where $\Theta^a={\cal E}\theta^a$. We now have to impose decoupling of unphysical modes in the perturbation theory. This can be accomplished by including in the nth order in the transformation rule a term of the form $g^n\theta (B^a)^n$, as was done for the QED case. The final transformation rule will be expressed in terms of nonlocal and field dependent representation operators ${\cal T}$ with a similar mixing between gauge and spacetime indices. As in the case of QED, the scattering amplitudes at the tree level will agree with the local Yang-Mills tree amplitudes. The regularized loop graphs will be finite to all orders as with the QED loop graphs. The non-Abelian action has the form:
\begin{equation}
S_{NA}=-\int d^4x\biggl[\frac{1}{4}B^{\mu\nu}B_{\mu\nu}+\bar\psi(i\slashed\partial-m)\psi\biggr]-\int d^4xd^4y\bar\psi(x){\cal V}[gB](x,y)\psi(y),
\end{equation}
where the vertex operator ${\cal V}[gB]$ is a spinorial matrix.

\section{Entire Functions and the S-matrix}

The nonlocal operator ${\cal E}(x)$ is defined by
\begin{equation}
{\cal E}(x-y)={\cal E}(x,\Lambda)\delta^{(4)}(x-y)={\cal E}(x,1/\ell^2)\delta^{(4)}(x-y),
\end{equation}
where $\ell$ is a constant with the dimensions of length. Because ${\cal E}$ is analytic (holomorphic) in the complex $z$ plane, we can expand it in a power series:
\begin{equation}
{\cal E}(x,\Lambda)=\sum_{n=0}^\infty\frac{c^n}{(2n)!}(Q(x,\Lambda))^n,
\end{equation}
where $Q$ is a specified operator. The operator ${\cal E}$ satisfies the following conditions:
\begin{enumerate}

{\item ${\cal E}(z)$ is an entire function of the order $1/2\leq\rho\leq 1$,

\item $[{\cal E}(z)]^*={\cal E}(z^*)$,

\item ${\cal E}(x) \geq 0\quad {\rm for\, real}\, x$,
\item ${\cal E}(0)=1,\quad {\cal E}'(0) <\infty.$

\item $\int_0^\infty dy{\cal E}(y) <\infty.$}
\end{enumerate}
The operator ${\cal E}$ is {\it not a physical field} with a corresponding pole in momentum space; it {\it does not introduce an extra physical degree of freedom}.

Let us consider an example of an entire function that can describe the operator ${\cal E}$~\cite{Boas,Knoppe,Krantz}:
\begin{equation}
\label{Airyfunction}
{\cal E}(x,\Lambda)=3^{2/3}\Gamma\biggl(\frac{2}{3}\biggr)Ai(\Box(x)/\Lambda^2),
\end{equation}
where $Ai(z)$ is the Airy function that is analytic in the entire finite complex plane ${\cal C}$ and ${\cal E}(0,\Lambda)=1$. The Airy function is a solution of the differential equation:
\begin{equation}
y''(z)-y(z)z=0,
\end{equation}
which has two linearly independent solutions:
\begin{equation}
y(z)=AAi(z)+BBi(z).
\end{equation}
Here, $A$ and $B$ are real constants and $Bi(z)$ is another Airy function. We have
\begin{equation}
Ai(z)=\frac{1}{z^{2/3}}\Gamma\biggl(\frac{2}{3}\biggr){}_0\!F_1\biggl(\frac{2}{3};\frac{1}{9}z^3\biggr)-\frac{z}{3^{1/3}}
\Gamma\biggl(\frac{1}{3}\biggr){}_0\!F_1\biggl(\frac{4}{3};\frac{1}{9}z^3\biggr),
\end{equation}
where $\Gamma$ is the gamma function and ${}_0\!F_1(z)$ is the confluent hypergeometric limit function. The Airy function has the series expansion:
\begin{equation}
\label{Airyseries}
Ai(z)=\frac{1}{3^{2/3}\pi}\sum_0^\infty\frac{\Gamma\biggl(\frac{1}{3}(n+1)\biggr)}{n!}\sin\biggl[\frac{2(n+1)\pi}{3}\biggr](3^{1/3}z)^n.
\end{equation}

For the special case $x > 0$ we obtain
\begin{equation}
{\cal E}(\Box/\Lambda^2)=\frac{N}{\pi\sqrt{3}}(\Box/\Lambda^2)^{1/2}K_{1/3}\biggl(\frac{2}{3}(\Box/\Lambda^2)^{3/2}\biggr),
\end{equation}
where $K_{1/3}(x)$ is a modified Bessel function of the second kind and $N$ is a normalization constant yielding ${\cal E}(0,\Lambda)=1$. In Euclidean momentum space, we get
\begin{equation}
{\cal E}(p_E^2/\Lambda^2)=\frac{N}{\pi\sqrt{3}}(p_E^2/\Lambda^2)^{1/2}K_{1/3}\biggl(\frac{2}{3}(p_E^2/\Lambda^2)^{3/2}\biggr).
\end{equation}
We have that ${\cal E}(p_E^2/\Lambda^2)\rightarrow 0$ in the limit $p_E^2\rightarrow\infty$. This follows from the asymptotic expansion of $Ai(z)$:
\begin{equation}
Ai(z)\propto \frac{1}{2\sqrt{\pi}(z)^{1/4}}\exp\biggl(-\frac{2}{3}z^{3/2}\biggr)\biggl[1-\frac{5}{48z^{3/2}}+\frac{385}{4608z^3}
+O\biggl(\frac{1}{z^{9/2}}\biggr)\biggr],
\end{equation}
where we have $\vert\,{\rm arg(z)}\vert < \pi$ and $\vert z\vert\rightarrow\infty$.

The S-matrix can formally be written in the interaction representation in the form of a T-product~\cite{Schweber}:
\begin{equation}
\label{Smatrix}
S=T\exp\biggl\{-i\int d^4x {\bar e}(x){\cal L}_I(x)\biggl\}
=T\exp\biggl\{-i\int d^4x {\bar e}(x){\bar\psi}(x)\gamma^\mu\psi(x)A_\mu\biggl\}.
\end{equation}
Here, the differential operator ${\bar e}(x)$ operates to the right on
\begin{equation}
\label{jcurrent}
j^\mu(x)={\bar\psi}(x)\gamma^\mu\psi(x).
\end{equation}
We expand the S-matrix in a series in $e$ and employ the Wick theorem on the $N$ product of the local field operators $\psi(x)$ and $A_\mu(x)$. The T-symbol is to be understood as a $T^*$ ordering constructed to make a meaningful perturbation theory for the interactions. The S-matrix becomes
\begin{equation}
S=\sum_{n=0}^n(-i)^n\frac{1}{n!}\int d^4x_1...\int d^4x_nT[{\bar e}(x_1){\cal L}_I(x_1)...{\bar e}(x_n){\cal L}_I(x_n)].
\end{equation}
In the application of the S-matrix, the interaction Lagrangian ${\cal L}_I$ will satisfy the microcausality condition:
\begin{equation}
[{\cal L}_I(x),{\cal L}_I(y)]=0\quad {\rm for}\quad (x-y)^2 < 0,
\end{equation}
where
\begin{equation}
{\cal L}_I(x)={\bar\psi}(x)\gamma^\mu\psi(x)A_\mu(x).
\end{equation}
The normal Wick ordering is performed on the physical local fields $\psi, {\bar\psi}$ and $A_\mu$.

The chronological contraction of the Dirac operator fields yields the usual causal propagator:
\begin{equation}
S(x-y)=< 0\vert T(\psi(x){\bar\psi}(y))\vert 0 >=\frac{i}{(2\pi)^4}\int\frac{d^4p\exp(-ip\cdot(x-y))}{{\slashed p}-m+i\epsilon}.
\end{equation}
The causal propagator for the photon field operator is
\begin{equation}
D_{\mu\nu}(x-y)=< 0\vert T(A_\mu(x)A_\nu(y))\vert 0 >=\frac{i\eta_{\mu\nu}}{(2\pi)^4}\int \frac{d^4k\exp(-ik\cdot(x-y))}{k^2+i\epsilon}.
\end{equation}

It is important to observe that the entire function operator ${\cal E}(x)$ occurring in the coupling ${\bar e}(x)=e{\cal E}(x)$ in the S-matrix (\ref{Smatrix}), described e.g., by the Airy function (\ref{Airyfunction}), {\it does not allow for a finite truncation of the series expansion} (\ref{Airyseries}). From the requirement that the entire function be of order $1/2\leq\rho\leq 1$, each term in the finite truncation of the series bears no resemblance to its analytic behavior (no singularity at infinity in the complex $z$ plane), or the asymptotic behavior of the exact description of the function. The entire function ${\cal E}(p^2)\rightarrow 0$ as $\vert p\vert^2\rightarrow\infty$, whereas its power series expansion has the property that every coefficient diverges as $\vert p\vert^2\rightarrow\infty$. This plays an important role if we require that we obtain a suitable $p^2$ dependence of ${\cal E}(p^2)$ in momentum space that leads to finite Feynman loop diagrams. The truncation of the series expansion and the result of the exact form of ${\cal E}(x)$ operating on ${\cal L}_I(x)$ do not commute. It is to be noted that the placement of the operator ${\bar e}(x)$ in (\ref{Smatrix}) plays an important role. We make the choice
\begin{equation}
S=T\exp\biggl\{-i\int d^4xA_\mu(x){\bar e}(x)j^\mu(x)\biggr\},
\end{equation}
where the operator ${\bar e}(x)$ acts on the current density $j^\mu={\bar\psi}(x)\gamma^\mu\psi(x)$.

\section{Self-Energy and Vacuum Polarization in QED}

Let us consider the self-energy of the electron corresponding to the term in the S-matrix:
\begin{equation}
-i:{\bar\psi}(x)\Sigma(x-y)\psi(y):,
\end{equation}
where
\begin{equation}
\Sigma(x-y)=-ie^2{\cal E}^2(x-y,\Lambda_{EM})R^\delta\gamma^\mu S(x-y)\gamma_\mu D(x-y).
\end{equation}
The regularizing function $R^\delta$ is defined by~\cite{Efimov,Moffat}:
\begin{equation}
R^\delta(z)=\exp[-\delta(z+iM^2)^{1/2+\nu}\exp(-i\pi\sigma)].
\end{equation}
Here, we have $0 < \nu <\sigma < 1/2$, and $M$ is a positive parameter. We have the estimates for $\vert z\vert\rightarrow\infty$:
\begin{equation}
\vert R^\delta\vert \sim \exp(-\delta\vert z\vert^{1/2+\nu})\quad {\rm for}\quad -\pi a_2 < {\rm arg}\,z < \pi(1+a_1),
\end{equation}
and
\begin{equation}
\vert R^\delta\vert \sim \exp(+\delta\vert z\vert^{1/2+\nu})\quad {\rm for}\quad \pi(1+ a_1) < {\rm arg}\,z < 2\pi(1-a_2/2),
\end{equation}
where
\begin{equation}
a_1=\frac{2(\sigma-\nu)}{1+2\nu},\quad a_2=\frac{1-2\sigma}{1+2\nu}.
\end{equation}
The function $R^\delta$ is analytic and decreases like an exponential function of order $\rho_1=1/2+\nu <1$ for $z$ in the upper half plane. The momentum integrals in the Feynman graphs will be convergent for $\delta > 0$. The limit $\delta\rightarrow 0$ in the integrals is taken after we have rotated the contours of integration over $p_0$ by an angle $\pi/2$. After rotating the argument $p_0\rightarrow ip_4$, the integrals are defined in 4-dimensional Euclidean momentum space. We require that ${\cal E}(p_E^2,\Lambda^2_{EM})\rightarrow 0$ as $p_E^2\rightarrow\infty$ where $p_E$ denotes the Euclidean momentum. Thus, the regularization procedure guarantees that we can analytically continue to the Euclidean metric when calculating Feynman loop diagrams.

We shall postulate that ${\cal E}(p^2,\Lambda^2_{EM}) \sim 1$ for $p^2 \lesssim \Lambda_{EM}^2$ where $\Lambda_{EM} > 1-2$ TeV. This guarantees that all our low energy calculations in our UV complete QED agree with currently available QED accelerator data.

In momentum space, we get
\begin{equation}
\Sigma(p)={\rm lim}_{\delta\rightarrow 0}(-ie)^2\int d^4x\exp(ip\cdot x){\cal E}^2(x,R^\delta)\gamma^\mu S(x)\gamma_\mu D(x)
$$ $$
={\rm lim}_{\delta\rightarrow 0}\frac{ie^2}{(2\pi)^4}\int d^4k{\cal E}(p-k,R^\delta)\gamma^\mu S(p-k){\cal E}(k^2,R^\delta)\gamma_\mu
D(k^2),
$$ $$
=\frac{e^2}{(2\pi)^4}\int d^4k_E{\cal E}(p_E-k_E)\gamma^\mu{\cal E}(k_E^2)\frac{m+{\slashed p}_E-{\slashed k}_E}{k_E^2(m^2+(p_E-k_E)^2)}\gamma_\mu.
\end{equation}
Here, we have
\begin{equation}
p_E=(ip_0, {\vec p}),\quad {\slashed k}_E=\gamma_0ik_4-{\vec\gamma}\cdot{\vec k},\quad p_E^2=-p^2.
\end{equation}
The self-energy of the fermion $\Sigma$ will be finite provided ${\cal E}(k_E^2)$ vanishes fast enough as $k_E^2\rightarrow\infty$.

We now consider vacuum polarization. The S-matrix contribution to first order is given by
\begin{equation}
-i:A^\mu(x)\Pi_{\mu\nu}(x-y)A^\nu(y):
\end{equation}
where
\begin{equation}
\Pi_{\mu\nu}(x-y)={\rm lim}_{\delta\rightarrow 0}(-ie^2){\cal E}^2(x,R^\delta)Tr[\gamma_\mu S(x-y)\gamma_\nu S(y-x)].
\end{equation}
Transforming to momentum space, we get
\begin{equation}
\Pi_{\mu\nu}(x-y)=\frac{1}{(2\pi)^4}\int d^4p\Pi_{\mu\nu}(p)\exp(ip\cdot (x-y)).
\end{equation}
We obtain in Euclidean momentum space:
\begin{equation}
\Pi_{\mu\nu}(p)=-i\frac{e^2}{(2\pi)^4}\int d^4k_E{\cal E}^2((k_E-p_E)^2)Tr\biggl[\gamma_\mu\frac{{\slashed k}_E-{\slashed p}_E-m}{(k_E-p_E)^2+m^2}
\gamma_\nu\frac{{\slashed k}_E-m}{k_E^2+m^2}\biggr].
\end{equation}
Evaluating ${\Pi^\mu}_\mu(0)$ we get
\begin{equation}
{\Pi^\mu}_\mu(0)= i\frac{2\alpha}{\pi^3}\int d^4k_E{\cal E}^2(k_E^2)\frac{p_E^2+2m^2}{(k_E^2+m^2)^2},
\end{equation}
where $\alpha=e^2/4\pi$ is the fine structure constant. To preserve gauge invariance we must require that ${\Pi^\mu}_\mu(0)=0$. For ${\cal E}(k_E^2)=1$, we see that ${\Pi^\mu}_\mu(0)$ is quadratically divergent and describes the standard QED result that the photon self-energy contribution violates gauge invariance. In our finite QED, provided ${\cal E}(k_E^2)$ vanishes sufficiently rapidly for $k_E^2\rightarrow\infty$, the calculation preserves gauge invariance and leads to a finite vacuum polarization. A detailed calculation of the QED vacuum polarization has been given in ref. [1]. For $\vert k\vert \lesssim \Lambda_{\rm EM}$ with $\Lambda_{EM}=1-2$ TeV, we recover the standard agreement with the Lamb shift experiment. From the vacuum polarization calculation, it can be shown that our UV complete QED does not possess a Landau pole.

In our perturbation theory formalism, the renormalization of mass and charge is finite in contrast to the standard local QED in which the bare mass and charge $m_0$ and $e_0$ and the self mass and charge $\delta m$ and $\delta e$ are infinite.

\section{The K\"all\'en-Lehmann Representation and Charges}

The K\"all\'en-Lehmann representation~\cite{Kallen,Lehmann} is obtained from the vacuum expectation values of two Heisenberg field operators. The basic assumptions are:
\begin{enumerate}

\item Relativistic invariance,

\item spectral conditions assuming the existence of a unique vacuum and only states with $p^\mu p_\mu \geq 0,\quad p_0 \geq 0$,

\item the physical states span a Hilbert space endowed with a Hermitian scalar product so that every state has a positive norm,

\item the physical field operators satisfy microcausality.
\end{enumerate}

We have translational invariance for QED:
\begin{equation}
<0\vert A_\mu(x)\vert {\vec p};p^2=m^2>=\exp(-ip\cdot x)<0\vert A_\mu(0)\vert {\vec p};p^2=m^2>=\frac{Z_3^{1/2}}{(2\pi)^{3/2}}\epsilon_\mu(p)\exp(-ip\cdot x),
\end{equation}
where  $A_\mu(x)$ is the Heisenberg photon field operator, $\epsilon_\mu$ is the photon field polarization vector and $Z_3$ is a real constant. The ``dressed" Feynman photon propagator $D'_{\mu\nu}(k^2)$ is given by
\begin{equation}
D'_{\mu\nu}(k^2)=\biggl(\eta_{\mu\nu}-\frac{k_\mu k_\nu}{k^2}\biggr)D'(k^2),
\end{equation}
where $D'(k^2)$ has the following spectral representation:
\begin{equation}
D'(k^2)=\frac{Z_3}{k^2}+\int_0^\infty dM^2\frac{\sigma(M^2)}{k^2-M^2+i\epsilon}.
\end{equation}
The constant $Z_3$ satisfies the condition
\begin{equation}
Z_3+\int_0^\infty dM^2\sigma(M^2)=1,
\end{equation}
and $0\leq Z_3\leq 1$. We obtain to second order in the coupling constant $e^2$:
\begin{equation}
Z_3^{(2)}\simeq 1-\frac{e^2}{12\pi^2}\ln\biggl(\frac{\Lambda_{EM}^2}{m^2}\biggr),
\end{equation}
with
\begin{equation}
\label{sigma2}
\sigma^{(2)}(M^2)=\frac{e^2}{12\pi^2}{\cal E}^2(M^2)\frac{1}{M^2}\biggl(1+\frac{2m^2}{M^2}\biggr)\biggl(1-\frac{4m^2}{M^2}\biggr)^{1/2}\theta(M^2-4m^2).
\end{equation}
The contribution $\sigma^{(2)}(M^2)$ comes from the $e^+e^-$ loop diagram. In the transverse gauge, we have
\begin{equation}
<0\vert T(A_\mu(x)A_\nu(y))\vert 0>\equiv D'_{\mu\nu}(x-y)=\biggl(\eta_{\mu\nu}-\frac{\partial_\mu\partial_\nu}{\Box}\biggr)D'(x-y).
\end{equation}
To order $e^2$ we get
\begin{equation}
\label{Dfunction}
D'_{\mu\nu}(x-y)=D_{\mu\nu}(x-y)+\frac{e^2}{8\pi}\int d^4u\int d^4z {\rm Tr}[\gamma^{\mu'}S(z-u)\gamma^{\nu'}S(u-z)]
$$ $$
\times {\cal E}(u)D_{\mu'\nu}(x-u){\cal E}(z)D_{\nu'\nu}(y-z),
\end{equation}
where for our choice of gauge:
\begin{equation}
D_{\mu\nu}(x-y)=\frac{2i}{(2\pi)^4}\int d^4k\exp(-ik\cdot(x-y))\biggl(\frac{k_\mu k_\nu}{k^2}-\eta_{\mu\nu}\biggr)\frac{i}{k^2+i\epsilon}.
\end{equation}

In contrast to QED with purely local interactions, (\ref{Dfunction}) is finite and the behavior of $D'_{\mu\nu}(k^2)$ depends on the large $k^2$ behavior of ${\cal E}(k^2)$.  We anticipate that we have
\begin{equation}
\label{klargelimit}
{\rm lim}_{k^2\rightarrow\infty}k^2D'(k^2)\sim Z_3\biggl[1+\int_0^\infty dM^2\sigma(M^2)+O\biggl(\frac{1}{k^2}\biggr)\biggr],
\end{equation}
as in QED with local point-like interactions.

We can obtain a physical interpretation of $D'(k^2)$ by considering the potential energy between two point charges with renormalized charges $e_{1R}=Z_3^{1/2}e_{10}$ and $e_{2R}=Z_3^{1/2}e_{20}$ separated by a distance $r$:
\begin{equation}
V(r)=\frac{e_{1R}e_{2R}}{(2\pi)^3}\int d^3k\exp(i{\vec k}\cdot{\vec r})D'(k^2).
\end{equation}
For large $r$ the potential $V(r)$ will be determined by small $k$ values with $\vert k\vert < \Lambda_{EM}=1-2$ TeV and we have
\begin{equation}
{\rm lim}_{r\rightarrow\infty}V(r)\sim \frac{e_{1R}e_{2R}}{4\pi r}+....
\end{equation}
The potential for $r\rightarrow 0$ is determined by the behavior of $D'(k^2)$ for $\vert k\vert > \Lambda_{EM}$:
\begin{equation}
{\rm lim}_{r\rightarrow 0}V(r)\sim e_{1R}e_{2R}U(r),
\end{equation}
where $U(r)$ is a finite function as $r\rightarrow 0$. Thus, in our UV complete QED the Coulomb potential singularity will be smeared out resulting in a finite value for $V(0)$.

The renormalized propagator $D'_{R\mu\nu}(k^2)$:
\begin{equation}
D'_{R\mu\nu}(k^2)=\biggl(\eta_{\mu\nu}-\frac{k_\mu k_\nu}{k^2}\biggr)D'_R(k^2),
\end{equation}
where $D'_R(k^2)$ satisfies the following spectral representation:
\begin{equation}
D'_R(k^2)=\frac{1}{k^2+i\epsilon}+\int_0^\infty dM^2\frac{\sigma_R(M^2)}{k^2-M^2+i\epsilon}.
\end{equation}
We also have that
\begin{equation}
Z_3^{-1}=1+\int_0^\infty dM^2\sigma_R(M^2).
\end{equation}
To lowest order in $e^2$, we now obtain from (\ref{sigma2}):
\begin{equation}
{\rm lim}_{M^2\rightarrow\infty}\sigma_R(M^2)\simeq \frac{e_R^2{\cal E}^2(M^2)}{12\pi^2}\frac{1}{M^2}.
\end{equation}
This leads to the result
\begin{equation}
Z_3^{-1}\simeq 1+\frac{1}{12\pi^2}\int^\infty\frac{dM^2}{M^2}e_R^2{\cal E}^2(M^2).
\end{equation}
Provided ${\cal E}(M^2)$ vanishes fast enough as $M^2\rightarrow\infty$, then $Z_3^{-1}$ is {\it finite} and $Z_3\neq 0$. For the standard local QED when ${\cal E}(M^2)=1$, we have $Z_3^{-1}=\infty$ to order $e^2$ and $Z_3=0$. This result implies that in QED with local interactions the ``bare'' charge $e_0^2=Z_3^{-1}e_R^2$ is infinite, or the renormalized charge $e_R^2$ vanishes. This can be interpreted, as before, that the Coulomb potential is singular at $r=0$. The coefficient of $1/r$ in the electrostatic potential between two charges is $Z_3^{-1}e_R^2$ at close distances in perturbation theory. In our UV complete QED the potential $V$ is smeared out as $r\rightarrow 0$. We also note that the requirement that $\sigma_R(M^2)$ is positive is met in our finite QED. We conclude from this that in our UV complete theory, we evade the ``triviality'' result in local QFT.

\section{High Energy Limit of Electromagnetic Form Factors}

The unsubtracted dispersion relations for electromagnetic form factors can be used for finite constant charge renormalization $Z_3^{-1}$, provided that the form factor for any vertex with two particles on the mass shell vanishes at infinite momentum. Lehmann, Symanzik, and Zimmermann~\cite{Zimmermann}, have shown that the vertex operator of QFT must satisfy a condition which implies that it vanishes at infinite momentum transfer, independent of any assumption about $Z_3^{-1}$.

Consider the matrix element for the reaction $e^++e^-\rightarrow\rightarrow e^++e^-$:
\begin{equation}
T=\frac{1}{(2\pi)^6}{\bar u}_e(k')({\bar e}(k-k')\gamma^\mu)v_e(k)\frac{\eta_{\mu\nu}}{(k-k')^2+i\epsilon}{\bar u}_e(p')({\bar e}((p-p')\gamma^\nu)v_p(p),
\end{equation}
where $S=1-T$. Because of the strong interactions of the quark with other particles, which also interact with the electromagnetic field, we write the more complicated $T$ for the reaction $e^++e^-\rightarrow\rightarrow q+{\bar q}$ in the form:
\begin{equation}
T=\frac{1}{(2\pi)^6}{\bar u}_e(k')({\bar e}(k-k')\gamma^\mu)v_e(k)\frac{\eta_{\mu\nu}}{(k-k')^2+i\epsilon}<\psi_{q'}\vert J_\mu(0)\vert\psi_q>.
\end{equation}
Here, we have
\begin{equation}
\label{Photonpotentialeq}
\Box A_\mu(x)=J_\mu(x)={\bar e}(x)j_\mu(x),
\end{equation}
and
\begin{equation}
(i\slashed\partial-m_q)\psi_q(x)=J_q(x),
\end{equation}
where $m_q$ is the quark mass, $A_\mu(x)$ is the renormalized Heisenberg electromagnetic field operator and
\begin{equation}
\label{Currentdensity}
J_\mu(x)={\bar e}(x){\bar\psi}_q(x)\gamma_\mu\psi_q(x).
\end{equation}

Let us consider the $q{\bar q}$ quark-photon form factors $F_1$ and $F_2$ defined by
\begin{equation}
<pp^{'(-)}\vert J_\mu(0)\vert 0>=\frac{1}{(4E_qE_{q'})^{1/2}}{\bar e}(q^2)<{\bar u}_q\vert[\gamma_\mu F_1(q^2)+\sigma_{\mu\nu}q^\nu F_2(q^2)]\vert v_{q'}>,
\end{equation}
where $q=p-p'$ is the virtual photon 4-momentum in the reaction $\gamma\rightarrow q+{\bar q}$, $p$ and $p'$ are the 4-momenta of the quark and antiquark, respectively; $E_q$ and $E_{q'}$ are the corresponding energies. Moreover, $<pp^{'(-)}\vert$ denotes an ingoing Heisenberg state of the $q{\bar q}$ pair, $\vert 0>$ is the Heisenberg vacuum state. The form of the spectral function for the photon propagator is~\cite{Drell}:
\begin{equation}
\sigma(q^2)=-\frac{1}{3}\sum_n(2\pi)^3\delta^{(3)}({\vec P}_n)2E_n\delta(E_n^2-q^2)\vert <n^{(-)}\vert A_\mu(0)\vert 0>\vert^2,
\end{equation}
and the polarization sum is positive definite. Consequently, it is bounded from below by the $q{\bar q}$ contribution alone. The latter state in the sum is
\begin{equation}
\sigma^{(q{\bar q})}(q^2)=\frac{1}{12\pi^2}\frac{1}{q^2}\biggl(1-\frac{14m_q^2}{q^2}\biggr)^{1/2}e^2{\cal E}^2(q^2/\Lambda^2_{\rm EM})\biggl[(F_1-4m_qF_2)^2
+\frac{2m_q^2}{q^2}\biggl(F_1-\frac{q^2}{m_q}F_2\biggr)^2\biggr].
\end{equation}

We now have that
\begin{equation}
Z_3^{-1}=1+\int dq^{'2}\sigma(q^{'2})\geq 1+\int dq^{'2}\sigma^{q{\bar q}}(q^2).
\end{equation}
Because $\sigma(q^2)$ is positive definite, it follows that if $Z_3^{-1}$ is finite, then $q^2\sigma(q^2)\rightarrow 0$ as $q^2\rightarrow\infty$ and it is necessary that ${\tilde F}_1\rightarrow 0$ and ${\tilde F}_2\rightarrow 0$ as $q^2\rightarrow\infty$ where
\begin{equation}
{\tilde F}_1(q^2)=e{\cal E}(q^2/\Lambda^2_{\rm EM})F_1(q^2),\quad {\tilde F}_2(q^2)=e{\cal E}(q^2/\Lambda^2_{\rm EM})F_2(q^2).
\end{equation}

Attempts to prove that at least one of the renormalization constants $Z_1=Z_2$ and $Z_3^{-1}$ is infinite have been made~\cite{Kallen2} without explicit use of perturbation theory. Assuming that QED is mathematically consistent and that the multiplicative renormalization constants are finite, K\"all\'en derived a formula for the asymptotic behavior of the renormalized vertex function $\Gamma_{R\mu}(p^2,p^{'2}, q^2)$ when the momentum transfer $q^2\rightarrow\infty$ and both $p$ and $p'$ are on the mass shell. K\"all\'en's result is that as $q^2\rightarrow\infty$:
\begin{equation}
e<pp^{'(-)}\vert j_\mu(0)\vert 0>\rightarrow \frac{1}{(4E_pE_{p'})^{1/2}}<{\bar u}_p\vert(e/Z_3)\gamma_\mu\vert v_{p'}>,
\end{equation}
where $j_\mu$ is given by (\ref{jcurrent}) and the reaction considered by K\"all\'en is $e^++e^-\rightarrow \gamma \rightarrow {\rm proton} + {\rm antiproton}$. This result would show that $F_1(q^2)\rightarrow e/Z_3$ as $q^2\rightarrow\infty$, implying that $Z_3^{-1}=0$ or $Z_3=\infty$. In our UV complete QED, this possible inconsistency of standard local QED is avoided provided ${\cal E}(q^2/\Lambda^2_{\rm EM})\rightarrow 0$ fast enough to guarantee that the nonlocal vertex function $\Gamma_{R\mu}\rightarrow 0$ as $q^2\rightarrow\infty$.

\section{$q{\bar q}$ Annihilation and QCD}

Finally let us consider the process $q+{\bar q}\rightarrow V^*\rightarrow \ell+\bar\ell$ where $V^*$ denotes a virtual $\gamma,Z, W$ and $\ell$ denotes a lepton. This is know as the Drell-Yan process~\cite{Drell2}. It has played an important role in determining the structure functions and analysis of the parton model in QCD~\cite{Halzen,Bodwin,Brodsky,Stasto}.

We begin with the parton sub-process cross section for $q+{\bar q}\rightarrow\gamma^*\rightarrow \ell^++\ell^-$:
\begin{equation}
\hat\sigma(q{\bar q}\rightarrow\ell^+\ell^-)=e^2_q\frac{4\pi\alpha^2(Q^2)}{3Q^2},
\end{equation}
where
\begin{equation}
Q^2=\hat{s}=(p_q+p_{{\bar q}})^2,
\end{equation}
is the $({\rm invariant\,mass})^2$, $e_q$ denotes the fractional charge of the quark and $\alpha(Q^2)=\alpha{\cal E}(Q^2/\Lambda_Q^2)$, where $\Lambda_Q$ is the nonlocal energy scale for the invariant mass $\sqrt{Q^2}$. We have
\begin{equation}
\label{Q4}
\frac{d\hat\sigma}{dQ^2}=e_q^2\frac{4\pi\alpha^2(Q^2)}{3Q^4}\delta(Q^2-\hat{s}).
\end{equation}
We obtain for the hadronic $pp$ cross section
\begin{equation}
\frac{d\sigma(pp\rightarrow\ell{\bar\ell}X)}{dQ^2}=\biggl(\frac{1}{3}\biggr)\sum_q\int dx_1\int dx_2f_q(x_1,Q^2)f_{{\bar q}}(x_2,Q^2)\frac{d\hat\sigma}{dQ^2},
\end{equation}
where we have summed over quark flavors. We have
\begin{equation}
\hat{s}=(x_1p_1+x_2p_2)^2\sim x_1x_2s
\end{equation}
and $s=2p_1\cdot p_2$ denotes the $({\rm center-of-mass\,energy})^2$ of the colliding protons. We now find that
\begin{equation}
\label{ppcrosssection}
\frac{d\sigma(pp\rightarrow \ell{\bar\ell}X)}{dQ^2}=\frac{4\pi\alpha^2(Q^2)}{9Q^4}\sum_qe_q^2\int dx_1\int dx_2f_{q}(x_1,Q^2)f_{\bar{q}}(x_2,Q^2)\delta\biggl(1-x_1x_2\frac{s}{Q^2}\biggr).
\end{equation}

From (\ref{Q4}) and (\ref{ppcrosssection}), we expect to obtain a scaling relation for minimal sub-quark interactions and gluon emission, provided that the parton distribution functions obey approximately:
\begin{equation}
f_q(x_1,Q^2)\sim f_q(x_1),\quad f_{{\bar q}}(x_2,Q^2)\sim f_{{\bar q}}(x_2).
\end{equation}
and $\alpha(Q^2)\sim\alpha$ for $\sqrt{Q^2}\lesssim\Lambda_Q$. The cross section is a function of the energy $\sqrt{s}$ and the lepton pair mass $\sqrt{Q^2}$ and
\begin{equation}
Q^4\frac{d\sigma}{dQ^2}=F\biggl(\frac{s}{Q^2}\biggr)
\end{equation}
is only a function of the ratio $s/Q^2$. The scaling relation is known to be satisfied from Fermi Laboratory data. The existence of a scaling law in the classical Drell-Yan process is dependent on the assumption of strong factorization~\cite{Bodwin,Brodsky,Stasto}. Provided $qq$ and $q$-gluon sub-processes are not important in the hadron production of lepton pairs, then approximate scaling relations can be expected to hold.

We have $\alpha(Q^2)\sim \alpha$ for $\sqrt{Q^2}\lesssim \Lambda_Q$, so that for low invariant mass energies $\sqrt{Q^2}$, we expect an approximate scaling relation to hold in our UV complete theory. However, for $\sqrt{Q^2}\gtrsim \Lambda_Q$, we predict a significant violation of the scaling relation, because of the $Q^2$ dependence of $\alpha(Q^2)$. The size of the scaling relation violation depends on how fast ${\cal E}(Q^2/\Lambda_Q^2)$ tends to zero for large $Q^2$ and on the measured size of $\Lambda_Q$. This prediction can be tested in proton-proton collisions at the LHC.

\section{Conclusions}

We have formulated UV complete QED and QCD theories that are gauge invariant to all orders in perturbation theory. The coupling constant in the action is promoted to an operator described by an entire function ${\cal E}(x,\Lambda)$, where $\Lambda$ is an energy scale that takes the constant value $\Lambda_{\rm EM}$ for QED and $\Lambda_{\rm NA}$ for non-Abelian gauge theory, respectively. The physical fields $\psi, {\bar \psi}$ and $A_\mu$ are described by local field operators that satisfy the condition of microscopic causality for spacelike separation $(x-y)^2 < 0$. In momentum space, ${\cal E}(p^2)$ does not possess a particle pole, so that the operator ${\cal E}(p^2)$ does not correspond to a physical field, but acts as a vertex function form factor. This guarantees that the scattering amplitudes satisfy unitarity and the Cutkosky rules~\cite{Cutkosky}. The S-matrix in the interaction representation provides a method for solving the equations in perturbation theory. The Feynman loop diagrams are finite to all orders of perturbation theory and the gauge invariance of the Lagrangian leads to the existence of Ward-Takahashi identities. The perturbative renormalizability of the mass and charge is finite, whereby the renormalization constants $Z_1,Z_2$ and $Z_3$ are finite to all orders.

We have performed a study of non-perturbative UV complete QED using the K\"all\'en-Lehmann representation and quark electromagnetic form factors. The long-standing issue as to whether the multiplicative renormalization constants $Z_1, Z_2=Z_1$ and $Z_3^{-1}$ are finite in QED is resolved by having the entire functions ${\cal E}(q^2/\Lambda^2_{\rm EM})$ vanish fast enough to guarantee that the quark form factors ${\tilde F}_{1,2}\rightarrow 0$ as $q^2\rightarrow\infty$.

For $\sqrt{s}\lesssim \Lambda$ where $\Lambda \geq 1$ TeV, the perturbative calculations lead to the standard results for QED and QCD, while for $\sqrt{s} > 1-2$ TeV our UV complete QFT will lead to new testable predictions that will differ in their high energy behavior from the standard local, renormalizable theory. An electroweak model without a scalar Higgs particle has been constructed that does not violate unitarity at $\sqrt{s}\gtrsim 1-2$ TeV~\cite{Moffat,Moffat2}. The model predicts scattering amplitudes and cross sections that will differ from the standard EW model including a Higgs particle. These predictions can be tested at the LHC.

\section*{Acknowledgements}

I thank Viktor Toth and Martin Green for stimulating and helpful conversations. This work was supported by the Natural Sciences and Engineering Research Council of Canada. Research at the Perimeter Institute for Theoretical Physics is supported by the Government of Canada through NSERC and by the Province of Ontario through the Ministry of Research and Innovation (MRI).

\end{document}